\documentclass[sigconf]{acmart}
\usepackage{multirow}
\AtBeginDocument{%
  \providecommand\BibTeX{{%
    Bib\TeX}}}

\copyrightyear{2023}
\acmYear{2023}
\setcopyright{rightsretained}
\acmConference[RecSys '23]{Seventeenth ACM Conference on Recommender Systems}{September 18--22, 2023}{Singapore, Singapore}
\acmBooktitle{Seventeenth ACM Conference on Recommender Systems (RecSys '23), September 18--22, 2023, Singapore, Singapore}\acmDOI{10.1145/3604915.3610644}
\acmISBN{979-8-4007-0241-9/23/09}

\begin{document}

\title{Turning Dross Into Gold Loss: is BERT4Rec really better than SASRec?}

\author{Anton Klenitskiy}
\email{antklen@gmail.com}
\affiliation{%
  \institution{Sber, AI Lab}
  \city{Moscow}
  \country{Russian Federation}
}

\author{Alexey Vasilev}
\email{alexxl.vasilev@yandex.ru}
\affiliation{%
  \institution{Sber, AI Lab}
  \city{Moscow}
  \country{Russian Federation}
}

\renewcommand{\shortauthors}{Klenitskiy and Vasilev}


\begin{abstract}


Recently sequential recommendations and next-item prediction task has become increasingly popular in the field of recommender systems. Currently, two state-of-the-art baselines are Transformer-based models SASRec and BERT4Rec. Over the past few years, there have been quite a few publications comparing these two algorithms and proposing new state-of-the-art models. In most of the publications, BERT4Rec achieves better performance than SASRec. But BERT4Rec uses cross-entropy over softmax for all items, while SASRec uses negative sampling and calculates binary cross-entropy loss for one positive and one negative item. In our work, we show that if both models are trained with the same loss, which is used by BERT4Rec, then SASRec will significantly outperform BERT4Rec both in terms of quality and training speed. In addition, we show that SASRec could be effectively trained with negative sampling and still outperform BERT4Rec, but the number of negative examples should be much larger than one.
\end{abstract}




\maketitle

\section{Introduction}
\label{sec:intro}

Recently Transformer-based models for sequential recommendations have received a lot of attention. There are two main approaches to using Transformers for sequential recommendations. The first one, which was introduced in SASRec paper \cite{kang2018self}, uses causal self-attention and only learns left-to-right relationships in a sequence. At training time at each time step model simply predicts the next interaction. Another way is to adopt a masked language modeling task which is widely used in Natural Language Processing. BERT4Rec \cite{sun2019bert4rec} uses a bidirectional Transformer to learn both left-to-right and right-to-left relationships. Some items in a sequence are randomly masked and the model tries to predict these masked items
by their surrounding context.

According to the original publication \cite{sun2019bert4rec} BERT4Rec achieved significant superiority over other deep learning approaches. Although there was some controversy regarding this question, most of the publications claim that BERT4Rec indeed achieves better performance than SASRec. These observations have been confirmed in a recent study on BERT4Rec replicability \cite{petrov2022systematic}.

However, the BERT4Rec training objective (predicting masked items) is only weakly related to the final goal of sequential recommendations. For SASRec, on the contrary, tasks for the stages of training and prediction are exactly the same - just predict the next item. Another disadvantage of the BERT4Rec approach is that it masks some of the items and calculates loss only for these masked items. For SASRec, on the other hand, all items in a sequence participate in loss calculations. As a result, SASRec gets more training signal from each training sequence than BERT4Rec. So we argue that it is not natural to expect a bidirectional model like BERT4Rec to be much better than a unidirectional model like SASRec.

Another difference apart from the training objective is that the two models use different loss functions. BERT4Rec applies softmax to the output layer and calculates cross-entropy over all possible items. In SASRec implementations for each positive item, only one negative item is sampled and binary cross-entropy loss is applied to these two items. So during backpropagation only weights for these two items get updates. This is a drastic contrast because for full cross-entropy loss weights for all items can get updates on each training step.

The main contributions of this work are: 
(1) We show that training the SASRec model with cross-entropy loss over all possible items makes it comparable to or even significantly superior to BERT4Rec. Moreover, unidirectional model training is much more efficient because BERT4Rec takes much more training time to get close to acceptable performance. This fact was already observed in works \cite{petrov2022systematic, petrov2022effective} ;
(2) We show that the SASRec model can be effectively trained without computing full loss over all items. It is possible to achieve good performance with negative sampling and cross-entropy loss, but the number of negative examples should be much larger than one. It is useful when the number of items in the catalog is large.

\section{Related Work}
\label{sec:literature}

Sequential recommender systems consider the order of interactions in the user's history. The goal of such systems is to predict the next item a user would be interested in. Early approaches to this problem used Markov Chains \cite{rendle2010factorizing, he2016fusing, he2017translation} for modeling sequential behavior. Later various deep learning models have been introduced, including recurrent (for example, GRU4Rec \cite{hidasi2015session, hidasi2018recurrent}) and convolutional (for instance, Caser \cite{tang2018personalized}) neural networks. After the arrival of the Transformer neural architecture \cite{vaswani2017attention} models based on the self-attention mechanism have been shown to achieve state-of-the-art performance and became prevalent.

Since the original SASRec \cite{kang2018self} and BERT4Rec \cite{sun2019bert4rec} papers a lot of research was done to further investigate the possibilities of Transformer-based models. Some works focused on improving self-attention mechanism (LSAN \cite{li2021lightweight}, LightSAN \cite{fan2021lighter}, Rec-denoiser \cite{chen2022denoising}), while others leverage additional side information (TiSASRec \cite{li2020time}, NOVA-BERT \cite{liu2021noninvasive}). Many publications introduced contrastive learning for sequential recommendations (CL4SRec \cite{xie2022contrastive}, CoSeRec \cite{liu2021contrastive}, DuoRec \cite{qiu2022contrastive}).

In order to accurately evaluate new state-of-the-art models, it is important to have good baselines. According to a recent study on BERT4Rec replicability \cite{petrov2022systematic}, in most of the publications, BERT4Rec outperforms the SASRec model. Also, it was shown that some papers used under-fitted versions of BERT4Rec, but with proper training, it can achieve performance comparable with newer algorithms. In our work, we address similar questions about whether it is possible to achieve good results with the original SASRec architecture.


\section{Loss functions}
\label{sec:loss}

Let's suppose that we have a set of users $U$ and a set of items $I$ with size $|I|$. Each user $u \in U$ is represented by his corresponding sequence of interactions with items $s_u=\{i^{(u)}_1, i^{(u)}_2,..,i^{(u)}_{n_u}\}$.

\begin{sloppypar}
Each sequential deep learning model (GRU4Rec, SASRec, BERT4Rec) acts as an encoder of input sequence $s_u$. The output of the last hidden layer is some representation of input sequence $H_u=SequenceEncoder(s_u)$, $H_u \in \mathbb{R}^{n_u \times d}$, where $d$ is hidden dimensionality of the model. It is used to calculate predicted relevances for items $R_u = H_uE^T$, where $E \in \mathbb{R}^{|I| \times d}$ is the item embedding matrix. Element $r_{ti}^{(u)}$ of matrix $R_u$ corresponds to the predicted relevance of item $i$ at time step $t$.
\end{sloppypar}
The original SASRec implementation doesn't make calculations with a full embedding matrix during training. Instead, it takes a true positive item, samples one negative item, and computes their relevances $r_{t,i_t}^{(u)}$ and $r_{t, -}^{(u)}$. Then for these two items, the classic binary cross-entropy loss is used:
\begin{equation}
    \label{loss:bce}
    \mathcal{L}_{BCE} = -\sum_{u \in U} \sum_{t=1}^{n_u} \log(\sigma (r_{t,i_t}^{(u)})) + \log(1-\sigma(r_{t,-}^{(u)})),
\end{equation}
where $\sigma()$ is the sigmoid function.

BERT4Rec implementations apply softmax over predicted relevances to get an output probability distribution for all items and compute cross-entropy loss:
\begin{equation}
    \label{loss:cross-entropy}
    \mathcal{L}_{CE} = -\sum_{u \in U} \sum_{t \in T_u} \log \frac{\exp(r_{t,i_t}^{(u)})}{\sum_{i \in I} \exp(r_{t,i}^{(u)})}
\end{equation}
where $r_{t,i_t}^{(u)}$ is predicted relevance for the ground truth item, and the second summation is done over a set of steps with masked items $T_u$. If we use this loss for unidirectional models (SASRec and GRU4Rec), summation will be done over all steps in a sequence: $T_u=\{1,2,..,n_u\}$.

The choice of the loss function is independent of the choice of the model architecture (GRU4Rec/SASRec/BERT4Rec) and training objective (item masking or next item prediction). Therefore, we propose to compare different models with the same loss. In section \ref{sec:results} we show, that if we train the SASRec model with cross-entropy loss as BERT4Rec, it achieves better performance and trains much faster. Hence, following the title of the paper, we propose to add more negative ("dross") items to the loss function to improve the quality of the models.

While training with cross-entropy loss over all items in the catalog leads to good performance, it can be computationally expensive or even unfeasible when the number of items becomes very large. To avoid this problem it is possible to sample  negative items for loss calculation. For each user sequence in a batch, we sample $N$ items a user hasn't interacted with and use the same set of negatives for each time step of a given sequence. As a result, we use the following sampled cross-entropy loss:
\begin{equation}
    \label{loss:cross-entropy_sampled}
    \mathcal{L}_{CE-sampled_N} = -\sum_{u \in U} \sum_{t=1}^{n_u} \log \frac{\exp(r_{t,i_t}^{(u)})}{\exp(r_{t,i_t}^{(u)}) + \sum_{i \in I^{-(u)}_N} \exp(r_{t,i}^{(u)})},
\end{equation}
where $I^{-(u)}_N$ is a set of $N$ negative examples sampled for a given user. This approach is computationally more efficient than sampling a separate set of negatives for each time step and leads to good performance when $N$ is large enough. A similar strategy for negative sampling was used in \cite{hidasi2018recurrent} to train the GRU4Rec model.

\section{Experimental Settings}
\label{sec:experimental_settings}
\subsection{Datasets}\label{sec:datasets}



We conduct experiments on five popular datasets, which are often used as sequential recommendations benchmarks. Amazon \textbf{Beauty} is a product review dataset crawled from Amazon.com \cite{mcauley2015image}. \textbf{Steam} is a dataset collected from Steam, a large online video game distribution platform \cite{pathak2017generating}. \textbf{MovieLens-1m} and \textbf{MovieLens-20m} are two versions of widely used movie recommendations dataset \cite{harper2015movielens}. \textbf{Yelp} is a business reviews dataset \cite{asghar2016yelp}. Unlike many previous publications, for exemple, \cite{xie2022contrastive,qiu2022contrastive,li2021lightweight}, we don't filter it by date and use the whole dataset to have more data and obtain more reliable evaluation results.

MovieLens-1m, MovieLens-20m, Amazon Beauty, and Steam have been used in original BERT4Rec publication \cite{sun2019bert4rec} and a recent study on BERT4Rec replicability \cite{petrov2022systematic}. For better reproducibility and fair comparison, we use exactly the same preprocessed versions of datasets from the BERT4Rec repository \cite{FeiSun}.

For all datasets, the presence of a review or rating was converted to implicit feedback, and users with less than 5 interactions were discarded. The final statistics of datasets are shown in Table \ref{tab:datasetStats}.

\begin{table}[ht]
\caption{Experimental datasets} 
\label{tab:datasetStats}
   \resizebox{0.5\textwidth}{!}{%
   \begin{tabular}{l r r r r r}
    \hline
    \textbf{Dataset} & \textbf{Users} & \textbf{Items} & \textbf{Interactions} & \textbf{Avg. len.} & \textbf{Density} \\ \hline
    ML-1M  &  $6,040$   & $3,416$  & $999,611$    & $165.49$ & $4.85\%$ \\\hline
    ML-20M &  $138,493$ & $26,744$ & $20,000,263$ & $144.41$ & $0.54\%$ \\\hline
    Steam  &  $281,428$ & $13,044$ & $3,488,885$  & $12.40$  & $0.10\%$ \\\hline
    Beauty &  $40,226$  & $54,542$ & $353,962$    & $8.79$ & $0.10\%$ \\\hline
    Yelp   &  $279,106$ & $148,415$& $4,350,510$   & $15.59$ & $0.11\%$ \\\hline
\end{tabular}
}
\end{table}

\subsection{Evaluation}\label{sec:evals}

To compare our results with previous works, we follow common practice \cite{kang2018self, sun2019bert4rec} and split each dataset into train, validation, and test partitions using the leave-one-out approach. For each user, the last item of the interaction sequence is used as the test data, the item before the last one is used as the validation data, and the remaining data is used for training.

In some previous publications, including original SASRec and BERT4Rec papers, sampled metrics were used for evaluation. For each positive item in the test set, 100 negative items are sampled, and only these items are used for metrics calculation. However, it was shown that sampled metrics can lead to inconsistent performance measures because they are not always consistent with unsampled metrics and depend on the sampling scheme and a number of negative examples \cite{krichene2020sampled, dallmann2021case, canamares2020target}. So we use full unsampled metrics for our experiments.

Performance is evaluated on two top-k ranking metrics, which are most widely used in other publications:  Normalized Discounted Cumulative Gain (NDCG@k) and Hit Rate (HR@k) with k={10, 100}. Note that for the leave-one-out strategy, HitRate is equivalent to another popular ranking metric - Recall. We take k=10 because it is the most popular value and is present in almost all publications. In previous works with sampling metrics, other popular values were k=5 and k=20. It was a reasonable choice because the ranking was made for 101 sampled items. But for full unsampled metrics and datasets with a large number of items, small values of k could not be very informative, so we chose k=100 as the second value.

\subsection{Models}\label{sec:models}

For a fair comparison, we train and evaluate all sequential models with the same code, which is present in our GitHub repository \cite{antklen}. We implement models with PyTorch and train them with the popular PyTorch Lightning framework \cite{Lightning-AI}.

We compare the following models in our experiments:

\textbf{BPR-MF} - a classic matrix factorization-based approach with a
pairwise BPR loss. We use fast GPU implementation of this model from
the Implicit library \cite{frederickson2018fast}.

\textbf{SASRec} - the original version of SASRec, which uses binary cross-entropy loss (\ref{loss:bce}). Code for model architecture was taken from the GitHub repository with the SASRec PyTorch implementation \cite{pmixer-AI} with slight adaptation to our training code.

\textbf{BERT4Rec} - BERT4Rec model. For the BERT backbone, we use the popular and efficient implementation from the HuggingFace Transformers library \cite{wolf2019huggingface, huggingface}.

\textbf{GRU4Rec} - our implementation of GRU4Rec model with cross-entropy loss (\ref{loss:cross-entropy}). We simply change the backbone from the Transformers model to the standard GRU layer remaining all other code is the same.

\textbf{SASRec+} - for short, we refer to our version as SASRec+. It is exactly the same model as the original SASRec but trained with cross-entropy loss (\ref{loss:cross-entropy}).

\textbf{SASRec+ <N>} - It is exactly the same model as the original SASRec, but trained with the sampled cross-entropy loss with $N$ negative items (\ref{loss:cross-entropy_sampled}).

\subsection{Implementation Details}\label{sec:impls}
For BPR-MF, we selected the best parameters (the number of latent components, regularization, and the number of iterations) with Optuna \cite{akiba2019optuna}. We trained the models with the learning rate 1e-3. The calculation was run 5 times for different seeds, and the metric values were averaged.

For all sequential models, we have tuned hidden size, number of self-attention blocks, and attention heads. For all models and all datasets except MovieLens-20M, we used a hidden size of 64. For the MovieLens-20M dataset, which is much bigger than others, a small hidden size leads to serious underfitting, so 256 was the best latent size. For SASRec, we used 2 self-attention blocks and 1 attention head. For BERT4Rec we used 2 self-attention blocks and 2 attention heads. The masking probability for BERT4Rec was set to 0.2. For MovieLens datasets that have a lot of long sequences, we set a maximum sequence length of 200. For all other datasets, we set a maximum sequence length of 50. All models were trained with a batch size of 128 and Adam optimizer with the learning rate 1e-3. These settings are consistent with parameters used in previous papers \cite{kang2018self, sun2019bert4rec, petrov2022systematic}.

To determine the number of training epochs, we use the early stopping criterion. We measure the NDCG@10 metric on the validation set and stop training if the validation metric does not improve for a given number of epochs (patience parameter). For SASRec and GRU4Rec models, we set patience to 10 epochs and a maximum number of epochs to 100. For BERT4Rec, we set a maximum number of training epochs to 200 and patience to 20 to be sure that the model is not underfitted because BERT4Rec needs more time to converge, as shown in section \ref{sec:speed}. After early stopping, we restore model weights from the best epoch on the validation set, this step could be important in some circumstances (see section \ref{sec:speed}). For datasets other than MovieLens-1m, we calculate validation metrics on a sample from the full validation set (we take 10000 random users) to speed up training.


\section{Results}
\label{sec:results}
\subsection{Overall Performance Comparison}\label{sec:comparison}

\begin{table*}[!ht]
\centering
\caption{Overall Performance Comparison. Training time is in seconds.} 
\label{tab:perf_comp}
\resizebox{0.78\textwidth}{!}{%
   \begin{tabular}{|l|l |r r |r r |r|r|}
    \hline
    \textbf{Dataset} & \textbf{Model} & \textbf{HR@10} & \textbf{HR@100} & \textbf{NDCG@10} & \textbf{NDCG@100} & \textbf{Training time} & \textbf{Best epoch} \\ \hline
    \multirow{2}{*}{ML-1M} & BPR-MF & $0.0762$ & $0.3656$ & $0.0383$ & $0.0936$ & $1$ & $60$ \\
    & GRU4Rec (our) & $0.2811$ & $0.6359$ & $0.1648$ & $0.2367$& $641$& $90$ \\
    & BERT4Rec & $0.2843$ & $0.6680$ & $0.1537$ & $0.2322$ & $1409$ & $197$ \\
    & SASRec & $0.2500$ & $0.6492$ & $0.1341$ & $0.2153$ & $486$ & $53$ \\
    & SASRec+ (our) & $\underline{0.3152}$ & $\underline{0.6743}$ & $\underline{0.1821}$ & $\underline{0.2555}$ & $540$ & $63$ \\
    & SASRec+ 3000 (our) & $\textbf{0.3159}$ & $\textbf{0.6808}$ & $\textbf{0.1857}$ & $\textbf{0.2603}$ & $769$ & $85$ \\\hline
    
    \multirow{2}{*}{ML-20M} & BPR-MF & $0.0806$ & $0.3373$ & $0.0394$ & $0.0892$ & $176$ & $350$\\
    & GRU4Rec (our) &$0.2813$ & $0.6153$ & $0.1730$ & $0.2401$& $6319$& $30$ \\
    & BERT4Rec &$0.2816$ & $0.6311$ &$0.1703$ & $0.2408$ & $14758$ & $68$\\
    & SASRec & $0.2001$ & $0.5932$ & $0.1067$ & $0.1851$ & $2495$ & $30$ \\
    & SASRec+ (our) & $\underline{0.2983}$ & $\underline{0.6397}$ & $\underline{0.1833}$ & $\underline{0.2521}$ & $9959$ & $46$ \\
    & SASRec+ 3000 (our) & $\textbf{0.3090}$ & $\textbf{0.6592}$ & $\textbf{0.1872}$ & $\textbf{0.2581}$ & $4125$ & $39$ \\\hline
    
    \multirow{2}{*}{Steam} & BPR-MF & $0.0431$ & $0.1767$ & $0.0223$ & $0.0480$ & $10$ & $370$ \\
    & GRU4Rec (our) & $0.1138$ & $0.3842$ & $0.0610$ & $0.1138$ & $869$ & $16$ \\
    & BERT4Rec & $\textbf{0.1242}$ & $\textbf{0.4132}$ & $\textbf{0.0662}$ & $\textbf{0.1228}$ & $4893$ & $74$ \\
    & SASRec & $0.0981$ & $0.3608$ & $0.0506$ & $0.1016$ & $2140$ & $39$ \\
    & SASRec+ (our) & $0.1191$ & $0.3947$ & $0.0641$ & $0.1179$ & $1262$ & $16$ \\
    & SASRec+ 3000 (our) & $\underline{0.1206}$ & $\underline{0.3974}$ & $\underline{0.0652}$ & $\underline{0.1192}$ & $1226$ & $14$ \\\hline
    
    \multirow{2}{*}{Beauty} & BPR-MF & $0.0271$ & $0.0970$ & $0.0144$ & $0.0280$ & $1$& $400$ \\
    & GRU4Rec (our) & $0.0291$ & $0.0933$ & $0.0163$ & $0.0288$ & $644$ & $25$ \\
    & BERT4Rec & $0.0338$ & $0.1051$ & $0.0187$ & $0.0325$ & $2325$ & $87$ \\
    & SASRec & $0.0246$ & $0.0939$ & $0.0126$ & $0.0262$ & $521$ & $26$ \\
    & SASRec+ (our) & $\textbf{0.0533}$ & $\textbf{0.1325}$ & $\textbf{0.0327}$ & $\textbf{0.0482}$ & $332$ & $6$ \\
    & SASRec+ 3000 (our) & $\underline{0.0490}$ & $\underline{0.1197}$ & $\underline{0.0295}$ & $\underline{0.0434}$ & $296$ & $8$ \\\hline

    \multirow{2}{*}{Yelp} & BPR-MF & $0.0176$ & $0.0967$ & $0.0087$ & $0.0236$ & $1$ & $100$ \\
    & GRU4Rec (our) & $0.0425$ & $0.1822$ & $0.0216$ & $0.0483$ & $2677$ & $5$ \\
    & BERT4Rec & $0.0442$ & $0.1912$ & $0.0223$ & $0.0505$ & $10166$ & $21$ \\
    & SASRec & $0.0228$ & $0.1067$ & $0.0115$ & $0.0274$ & $655$ & $3$ \\
    & SASRec+ (our) & $\textbf{0.0482}$ & $\textbf{0.2005}$ & $\textbf{0.0246}$ & $\textbf{0.0539}$ & $2505$ & $3$ \\
    & SASRec+ 3000 (our) & $\underline{0.0462}$ & $\underline{0.1929}$ & $\underline{0.0237}$ & $\underline{0.0519}$ & $1965$ & $5$ \\\hline
    
\end{tabular}
}

\end{table*}

\begin{table*}[ht!]
\centering
\caption{Results reported
in the literature for the MovieLens-1m dataset. Metrics are
copied from the respective publications.} 
\label{tab:oldworks}
\resizebox{0.8\textwidth}{!}{%
   \begin{tabular}{|l|r|r|l|r|}
    \hline
    \textbf{Publication} & \textbf{SASRec NDCG@10} & \textbf{BERT4Rec NDCG@10}& \textbf{Best model} & \textbf{Best model NDCG@10} \\\hline
    This paper & 0.1341 & 0.1537 & SASRec+ 3000 & 0.1857 \\\hline
    Petrov et al. \cite{petrov2022systematic} & $0.1078$ & $0.1516$ & ALBERT4Rec & $0.165$ \\\hline
    Du et al. \cite{du2022contrastive} & $0.0918$ & 0.1097 & CBiT & $0.1694$ \\\hline
    Fan et al. \cite{fan2021lighter} & $0.1121$ & $0.1099$ & LightSANs & $0.1151 $ \\\hline
    Qiu et al. \cite{qiu2022contrastive} & $0.0910$ & $0.0619$ & DuoRec & $0.168$ \\\hline
    Liu et al. \cite{liu2021noninvasive} & - & $0.1398$ & NOVA-BERT & $0.168$ \\\hline
    \end{tabular}
}

\end{table*}

Table \ref{tab:perf_comp} summarizes the results of experiments on all five datasets. For sampled cross-entropy loss (\ref{loss:cross-entropy_sampled}) we show metrics for $N=3000$. Performance for other values of $N$ is analyzed in section \ref{sec:sampling_train}.

Our experiments confirm previous results \cite{petrov2022systematic} that BERT4Rec is persistently better than vanilla SASRec with binary cross-entropy loss (\ref{loss:bce}). However, when we train SASRec with cross-entropy loss (\ref{loss:cross-entropy}) or (\ref{loss:cross-entropy_sampled}), the situation is reversed. On all datasets except Steam SASRec+ and SASRec+ 3000 significantly outperform BERT4Rec. Moreover, BERT4Rec needs much more training time to achieve moderate performance.

Remarkable that for MovieLens datasets even a good old GRU4Rec baseline could be competitive with BERT4Rec. This observation supports our opinion that unidirectional causal modeling is more appropriate for the next item prediction task than the bidirectional masking approach from BERT.

In Table \ref{tab:oldworks} we compare our results with previous works which used the same unsampled metrics on the MovieLens-1m dataset. The performance of our BERT4ec implementation is comparable with other papers, so our version is not underfitted. As for vanilla SASRec implementation, our numbers are better because we used longer maximum sequence lengths. If we train our vanilla SASRec model with a maximum sequence length of 50, as in other works, NDCG@10 will be equal to 0.1135. This value is pretty close to some other publications.

\subsection{Convergence speed}\label{sec:speed}

\begin{figure*}
\includegraphics[width=\textwidth,height=3cm]{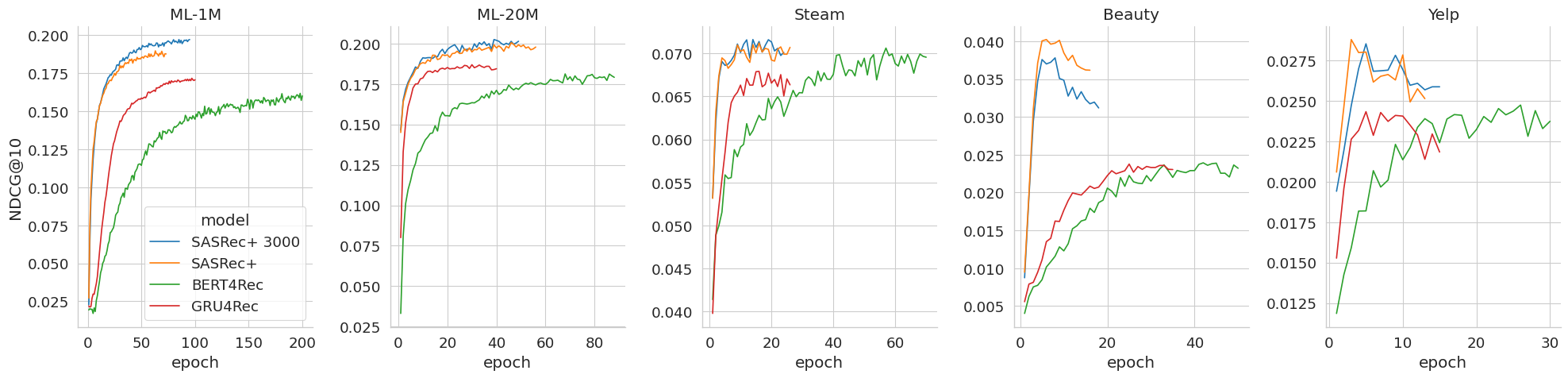}
\caption{Convergence speed of different models. Blue line corresponds to SASRec+ 3000, orange line - SASRec+, red line - GRU4Rec, green line - BERT4Rec.}
\label{fig:convergence_speed}
\Description[Convergence speed of different models]{Convergence speed of different models is in epochs number}
\end{figure*}

To better analyze the convergence speed of different models, we plot the NDCG@10 metric on the validation set against the epoch number. Figure \ref{fig:convergence_speed} demonstrates such curves for GRU4Rec, BERT4Rec, and our versions of SASRec on all datasets. It is clear that BERT4Rec needs much more training time and epochs to achieve satisfactory performance. This observation is consistent with recent works \cite{petrov2022systematic, petrov2022effective}.

It is worth noting that on Beauty and Yelp datasets, SASRec learns very quickly but then starts to overfit and validation performance degrades. BERT4Rec on the other hand doesn't overfit and oscillates near the final performance level. So in some circumstances, it is important to restore the best model weights after early stopping to achieve the best possible performance.

\subsection{Training with negative sampling}\label{sec:sampling_train}

\begin{figure*}
\includegraphics[width=\textwidth,height=3cm]{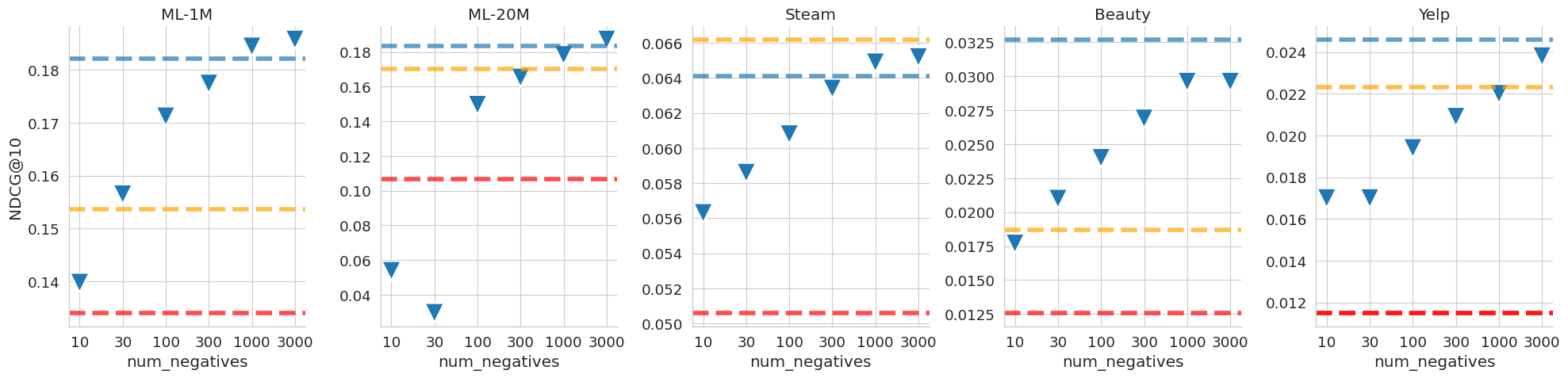}
\caption{NDCG@10 for a different number of negative examples. Markers represent performance for a corresponding number of negatives. SASRec+ (blue line), SASRec (red line), and BERT4Rec (orange line) are added for comparison.}
\label{fig:num_negatives}
\Description[NDCG@10 for a different number of negative examples.]{NDCG@10 for a different number of negative examples. With the growth of negative examples, the quality grows.}
\end{figure*}

Figure \ref{fig:num_negatives} demonstrates the performance for different numbers of negatives in sampled cross-entropy loss (\ref{loss:cross-entropy_sampled}). It starts from modest values for small $N$ and achieves or almost achieves the performance of SASRec+ with full cross-entropy loss (\ref{loss:cross-entropy}) for a large number of negative items. We conclude that training with $N \approx 1000$ is a good option, though the appropriate value of $N$ could depend on the dataset at hand.

\section{Conclusion}
\label{sec:conclusion}

In this work, we show that with proper training unidirectional SASRec model is still a strong baseline for sequential recommendations. Previous works used binary cross-entropy loss with one negative example. If trained with the cross-entropy loss on a full item set or sampled cross-entropy loss with a large number of negative examples, it can outperform the bidirectional BERT4Rec model and achieve performance comparable with current state-of-the-art approaches. We encourage to use SASRec with the cross-entropy loss for future research as a baseline for more rigorous evaluation of new state-of-the-art algorithms.

\bibliographystyle{ACM-Reference-Format}
\bibliography{seq}

\end{document}